\documentclass[aps,twocolumn,pra,showpacs]{revtex4}

\usepackage{graphicx}
\usepackage{amssymb}
\usepackage{color}%
\usepackage{amsmath}%
\usepackage{dcolumn}% Align table columns on decimal point
\usepackage{bm}% bold math
\usepackage[dvipdfm]{xcolor}
\usepackage{subfigure}

\usepackage{amsfonts}
\begin{document}
\title{Mode competition in superradiant scattering of matter waves}
\author{Thibault Vogt}
\email{thibault.vogt@pku.edu.cn}
\author{Bo Lu}
\author{XinXing Liu}
\author{Xu Xu}
\author{Xiaoji Zhou}
\email{xjzhou@pku.edu.cn}
\author{Xuzong Chen}
\email{xuzongchen@pku.edu.cn}
\affiliation{School of Electronics Engineering and Computer Science, Peking University, Beijing 100871, People's Republic of China}
\date{15 May 2010}

\begin{abstract}

Superradiant Rayleigh scattering in a Bose gas released from an optical lattice is analyzed with incident light pumping at the Bragg angle for resonant light diffraction.
We show competition between superradiance scattering into the Bragg mode and into end-fire modes clearly leads to suppression of the latter at even relatively low lattice depths. A quantum light-matter interaction model is proposed for qualitatively explaining this result.

\end{abstract}
\maketitle

The coherent nature of Bose-Einstein condensates has led to new and rapid developments in atom optics and studies on coherent interaction between light and matter waves, with the demonstration of efficient matter wave interferometers, Bragg diffraction, wave mixing, matter wave amplifiers. Superradiant scattering was for the first time analyzed using a Bose-Einstein condensate (BEC) in a seminal experiment by Ketterle et al. \cite{Inouye}. In this experiment the initial matter grating, formed due to Rayleigh scattering of a pumping beam by an elongated Bose-Einstein condensate and subsequent recoil into a moving matter wave, was self amplified by resonant light diffraction in a phenomenon called amplification of matter waves. Absorption of pumping photons and preferential scattering into so-called end-fire modes along the BEC's long axis lead to the observation of patterns of coherently recoiling atoms.
Amplification of matter waves was further characterized with the use of an initial matter wave seed formed via Bragg diffraction of a Bose condensate and its coherence nature was demonstrated \cite{inouye1999phase,kozuma1999phase}.

For an elongated BEC, superradiant Rayleigh scattering light is emitted along the long axis because the gain for superradiance is maximum in this direction. However, different superradiant modes can be obtained when a matter wave seed with non-zero momentum is created before pumping since light can then be resonantly diffracted in a different direction \cite{Inouye2}.
No precise analysis of competition between different superradiant scattering modes has been carried out yet. Such analysis would be useful for calibrating precisely superradiance gains and initial Rayleigh scattering rates if one wants to use quantitatively superradiance for the analysis of coherence in Bose gases \cite{sadler,PhysRevLett.97.180410}. Analyzing superradiance with non common configurations is also important as combining early stage superradiance described with quantum theory and long timescale superradiance which is well captured within a semi-classical theory taking into account propagation effects is currently a topic of high interest \cite{PhysRevLett.105.220404,PhysRevA.82.023608}. 

\begin{figure}[hptb]
\begin{center}\includegraphics[width=7cm] {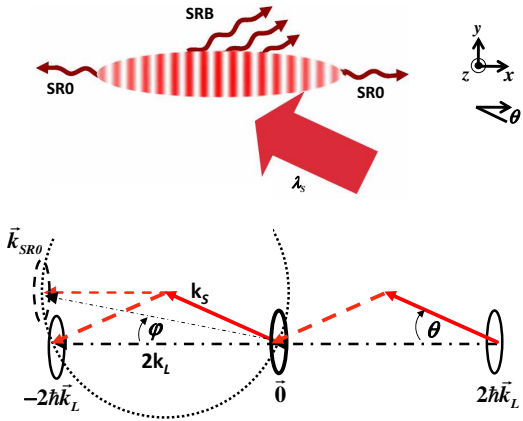}
\caption{(Color online) (top) Scheme of the superradiance experiment. A pump beam (SR) is sent at $\theta=24^{\circ}$ with respect to the long axis of a BEC after its loading in a 1D optical lattice. This beam ($\lambda_S =780$~nm) can be superradiantly scattered into three modes, a Bragg mode (SRB) symmetric of the incident beam with respect to the $x$ axis and two end-fire modes along the $x$ axis (SR0). (Bottom) Momentum representation of the Bose gas during the superradiance process. Absorption of photons by the condensate at rest (momentum $\vec 0$) followed by emission at $24^{\circ}$ into the Bragg mode leads to amplification of the matter wave (SRB) with momentum $-2 \hbar \vec k_L$. One superradiant mode ($\vec k_{SR0}$) recoiled at $\varphi=\theta/2$ with respect to the $x$ axis is also shown. It corresponds to absorption of one pump photon and subsequent emission into the right scattered end-fire mode shown on the top picture. The dotted circle stands for the energy conservation for atoms initially at rest during the Rayleigh scattering process.}%
\label{setup}%
\end{center}
\end{figure}

We present in this article an experimental analysis of mode competition in superradiance scattering. 
Rather than relying on an initial matter wave seed formed via Bragg diffraction of a BEC \cite{inouye1999phase,kozuma1999phase,Inouye2} our superradiance experiment is performed after initial adiabatic loading of a BEC along its long axis inside a 1D optical lattice \cite{Xuxu}. The choice of contra propagating optical beams (wave vector $\pm \vec k_L$) for the optical lattice loading leads to the formation of a matter grating with period $d=\pi/k_L$, so that three waves are mostly populated after release from the optical lattice (see Fig. \ref{setup}). 
In the superfluide regime, the one-particle wave function of the system just after lattice switch off is indeed $\psi(\vec r) \approx \phi_0(\vec r) \left(\tilde{w}(0) + \tilde{w}(-2 \vec k_L) e^{2ik_L x} + \tilde{w}( 2 \vec k_L) e^{-2ik_L x} \right)$, where $\tilde{w}$ designates the Fourier transform of the Wannier function $w(x)$ whose width has been neglected compared to the width of $\phi_0(\vec r)$ the profile of the Bose gas loaded in the optical lattice \cite{krämer2002macroscopic}. The two recoiling waves with amplitudes proportional to $\tilde{w}(-2 \vec k_L)=\tilde{w}(2 \vec k_L)$ are only lightly populated and one, for example with momentum $-2 \hbar \vec k_{L}$, can be used as seed wave in a superradiance experiment.
The angle and frequency of a pump pulse for superradiance has to be chosen at the corresponding Bragg angle for optical diffraction, which is the condition for amplification of this seed wave \cite{Inouye2}. Indeed, with the absorption of a pump photon by an atom and emission into the resonantly diffracted mode, with total momentum change $- 2\hbar \vec k_{L}$, the amplitude of the matter wave with momentum $- 2 \hbar \vec k_{L}$ as well as the diffraction efficiency are further increased (see Fig. \ref{setup}).
Besides this superradiance light scattering into the Bragg mode (further named SRB), usual superradiance light scattering into end-fire modes along the long axis (further named SR0) can occur as well.
We study in the following mode competition between SRB and SR0. It is shown that SRB is clearly enhanced at even relatively low lattice depths whereas SR0 is suppressed. Those results are used for attempting to calibrate the superradiance gains obtained in quantum theory of supperadiance \cite{HanPhysRevLett91150407}.

{\it Description of the experiment.} ---Our experiment (see Fig.\ref{setup}) is carried out with a BEC of $~2.5\times 10^{5}$ $^{87}$Rb atoms in the $F=2, m_{F}=2$ state \cite{YangFan,zhou}.
Obtained after trapping of cold atoms in a QUIC trap (longitudinal trapping frequency $\omega_x= 20~$Hz and radial trapping frequency $\omega_r=200$~Hz) and further radio frequency evaporative cooling, our BEC is cigar-shaped, with $70\ \mu m$ length and $7\ \mu m$ width.

The BEC is loaded along its long axis in an optical lattice formed with a retro-reflected laser beam ($\lambda_{L}=852$ nm), focused on the BEC to a waist of $110\ \mu$m. For the system to remain in the superfluide state, we load the atoms with a 40~ms exponential rising of the optical lattice depth up to a desired value $V_0$ lower than $15\ E_{r}$, $E_{r}=\hbar^{2}k_{L}^2/2m$ being the recoil energy of a Rubidium atom. We then keep the optical lattice switched on for $50$ ms before sudden release of the combined optical and magnetic traps.

Just after release, a $30\ \mu$s light pulse, detuned from the D2 transition ($\lambda_{S} =780$ nm) by 1.45~GHz with respect to the excited $F'=3$ state, is sent onto the matter wave grating, with polarization perpendicular to the BEC long axis. The chosen angle $\theta \approx 24^{\circ}$ between the incident light direction and the long axis of the condensate satisfies the condition $\cos\theta=\lambda_{s} / \lambda_{L}$. 
As shown in Fig. \ref{setup}, the pump pulse can be resonantly diffracted by the matter wave grating at $24^{\circ}$ in a direction symmetric of the incident beam one with respect to the long axis of the BEC.
Eventually, absorption imaging is performed after $30$ ms time of flight along the $z$ axis.

\begin{figure}[hptb]
\begin{center}
\includegraphics*[angle=0,width=7.8cm]{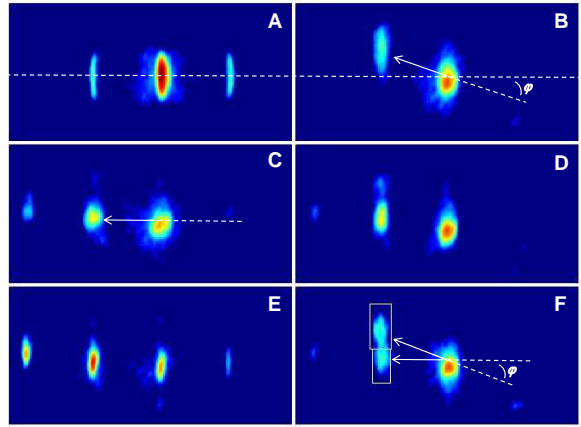}
\caption{(Color online) 30 ms time of flight absorption images. (A) Multiple matter-wave interference patterns after atoms were released from an optical lattice potential with a potential depth of $14.4\ E_{r}$. (B) SR0 after a laser pump pulse is sent on the BEC. (C) SRB when a pump pulse is sent on a Bose gas previously loaded in an optical lattice at $4.5\ E_{r}$ depth. (D) SR0 and SRB with the same pump pulse and at $2.7\ E_{r}$ optical lattice depth. (E) SRB after previous loading in an optical lattice at $14.4\ E_{r}$ optical depth. (F) SR0 and SRB for at $1.8\ E_{r}$ lattice depth and higher pump intensity (see text). The dynamic range is divided by a factor of 2 for this last image. The superradiance driving pulses are 30 $\mu$s long.}%
\label{tof}%
\end{center}
\end{figure}

{\it Mode competition between SRB and SR0}---
Given the choice of angle, scattered  light amplification at $24^{\circ}$ and simultaneous matter wave amplification (SRB) can occur in the presence of the density grating resulting from the optical lattice. Usual Rayleigh scattering superradiance (SR0) with scattering into end-fire modes can also be observed for low lattice depths. We study here competition between those two processes.

\begin{figure}
[hptb]
\begin{center}
\includegraphics*[
width=6.5cm,height=4.0cm]{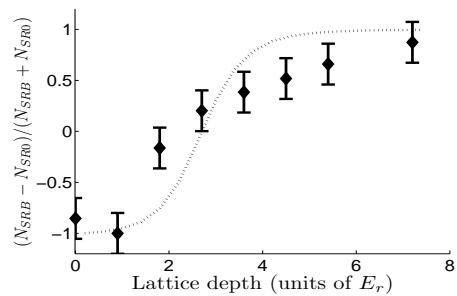}%
\caption{ Comparison of the numbers $N_{SR0}$ of atoms scattered in the forward SR0 mode (see text) and $N_{SRB}$ in the SRB mode, for different lattice depths. The dotted line is a corresponding simulation of superradiance scattering.}%
\label{threshold}%
\end{center}
\end{figure}

Different absorption images are shown in Fig. \ref{tof}, corresponding to pump pulses of 30 $\mu$s and intensity $95$~mW/cm$^2$ from Fig.  \ref{tof} (B) to (E).
In (B), the SR0 recoiling pattern is shown when the pump is shined directly onto the BEC without initial loading in the optical lattice.
Essentially one forward mode can be observed, corresponding to a recoil at $\theta  /2=12^{\circ}$ with momentum $k_{SR0}=2 k_{S}\cos\left(\theta  /2 \right)$. It corresponds to the absorption of one photon from the pump with emission into an end-fire mode. Interestingly, the recoiling pattern does not show doublets seen when pumping at $90^{\circ}$ and corresponding to the presence of both end-fire modes. They can actually be seen at higher pump intensities, suggesting that one end-fire mode is preferentially amplified.
In (C), when sending the same pump pulse but with initial loading of the BEC inside the optical lattice, a very strong amplification of the $-2\hbar \vec k_{L}$ order is observed, while the  0 and $+2\hbar \vec k_{L}$ orders are depleted. This is the confirmation of SRB. As a result of mode competition, SR0 is suppressed in the presence of the optical lattice, leaving only SRB. We find that SR0 is completely suppressed when the lattice depths are higher than $~4.5 E_{r}$, as it appears in Fig. \ref{threshold}, where we compare the numbers of atoms in the first SR0 and SRB scattering orders. A corresponding absorption image at $2.7 E_r$, where both SR0 and SRB are present, is shown in Fig. \ref{tof} (D). In (E), complete suppression of SR0 is confirmed and strong amplification of higher order modes appears as well for a 14.4 $E_r$ lattice depth. The upper part of the mode at rest seems to be more affected by the amplification process, which might be due to stronger localization of the Bragg diffracted beam on the top part of the BEC. In (F), a picture taken for similar pump pulse duration but higher intensity, $160$~mW/cm$^2$, illustrates also mode competition between SRB and SR0.
In the presence of the optical lattice grating, SRB shows a nearly $2.6^{\circ}$ half maximum resonance gain width around $\theta$, while far from this position, SR0 is no more suppressed at low lattice depths and shows only little dependence on the angle. 
Another interesting factor is the detuning of the pump beam. Whatever the lattice depth is, SR0 is not observed in the case of a blue detuning because of a propagation effect of the end-fire modes inside the matter medium, generating an optical-dipole potential dependent on the sign of the detuning \cite{Deng}. In the presence of the lattice, SRB is as strong for red or blue detuning, which shows that this propagation effect is much weaker or absent in this case. This result could be due to the fact light travels the condensate on a much shorter distance and then leaves it in a much shorter time than when the superradiant light propagates along the long axis of the BEC. Analysis of this result is still under consideration.

For simulating our data, it is important to be explicit on how the number of atoms $N_{SR0}$ and $N_{SRB}$ are measured in Fig. \ref{threshold} since a small overlap occurs between SR0 and SRB modes in time of flight absorption imaging.
Actually, in the areas delimited by the white rectangles in Fig. \ref{tof} (F), we calculate the spatially integrated numbers of atoms $N_1$ and $N_2$ that are given by:
\begin{align}
N_1=aN_{SRB}+bN_{SR0}\notag\\
N_2=a'N_{SR0}+b'N_{SRB}\notag
\end{align}
with $a'+b=1$ and $a+b'=1$. From this, we compute $N_{SRB}=\left(a'N_1-(1-a')N_2\right)/\left(a+a'-1\right)$ and $N_{SR0}=\left(aN_2-(1-a)N_1\right)/\left(a+a'-1\right)$. With our choice of areas of integration, slightly different values of $a'$ and $a$ of $\sim 0.2-0.3$ are obtained from time of flight measurements at $0\ E_r$ and more than $8\ E_r$ respectively. For the particular choice of the areas of integration where $a \approx a'$, $(N_{SRB}-N_{SR0})/(N_{SRB}+N_{SR0})$ is directly proportional to $(N_{1}-N{2})/(N_{1}+N{2})$.

Interpretation of the data is carried out using the quantum model of references \cite{Inouye,Moore,HanPhysRevLett91150407} adjusted to take into account the seeding. 
The matter wave field is given by $\Psi(\vec r,t)=\sum_{\vec q} \phi_{\vec q}(\vec r) e^{i \vec q \vec r} \hat{c}_{\vec q}(t)$, limited to the first scattering orders $\hat{c}_0$, $\hat{c}_{\vec k_{SR0}}$, $\hat{c}_{2\vec k_L}$ and $\hat{c}_{-2\vec k_L}$, the annihilation mode operators of atoms with corresponding momenta and profiles $\phi_{\vec q}(\vec r)$. The form of the initial wave function implies $\phi_{\vec q}(\vec r)=\phi_{\vec 0}(\vec r)$. The time evolution of the mode operators is given by~\cite{HanPhysRevLett91150407}:

\begin{equation}
\begin{split}
\dot{\hat{c}}_{ \vec k_{SR0}}&= -i \omega_{ \vec k_{SR0}}\hat{c}_{ \vec k_{SR0}}+\frac{G_{ \vec k_{SR0}}}{2} N_0 \hat{c}_{ \vec k_{SR0}}+\hat{f}^\dagger_{ \vec k_{SR0}} \\
\dot{\hat{c}}_{\pm 2 \vec k_L}&= -i \omega_{ 2 \vec k_L}\hat{c}_{\pm 2 \vec k_L} \mp \frac{G_{-2 \vec k_{L}}}{2} N_0 (\hat{c}_{\pm 2 \vec k_L}+\hat{c}^{\dagger}_{\mp 2 \vec k_L})\label{rate}
\end{split}
\end{equation}

\noindent In this expression, $G_{ \vec k_{SR0}}$ and $G_{-2 \vec k_{L}}$ are the SR0 and SRB gains. $\omega_{ \vec k_{SR0}}$ and $\omega_{ 2 \vec k_L}$ are the recoil frequencies of the quasi modes. The operator $\hat{f}^\dagger_{ \vec k_{SR0}}$ represents the initial scattering into mode $\hbar \vec k_{SR0}$ due the interaction of the Bose gas with vacuum. $N_0$ is the number of atoms in the condensate at rest. The superradiance gains are of the form $G_{\vec q} = \frac{2 \pi}{c} \int d \vec k |g(\vec k)|^2 |\rho_{\vec q}[\vec k,t]|^2\delta[|\vec k|-|\vec k_S|]$. They depend on the integral $\rho_{\vec q}[\vec k,t] = \int d \vec r |\phi_0(\vec r,t)|^2 \exp[-i(\vec k-\vec k_S + \vec q)\cdot \vec r]$, centered at $\vec k_S-\vec q$, and on $g(\vec k)=\frac{|\Omega_0 \mu|}{2|\delta|}(\frac{c k_S}{2\hbar\varepsilon_{0}(2\pi)^{3}})^{1/2}|\vec k\times \vec z|$
the atomic coupling coefficient between pump and scattered light, $\mu$ and $\Omega_0$ referring, respectively, to the dipole moment of the transition involved and its Rabi frequency. 
The SRB gain is related to $G_{ \vec k_{SR0}}$ the gain for the SR0 process by the relation $ G_{SRB} = G_{-2 \vec k_{L}}= G_{ \vec k_{SR0}}/ \sqrt{\cos^2 \theta + (L/W)^2 \sin^2 \theta}\approx 0.22 G_{ \vec k_{SR0}}$ \cite{Moore}, with $L$ and $W$ the length and width of the Bose gas after its loading inside the optical lattice.
In the superfluid phase, the Bose gas conserves its aspect ratio \cite{krämer2002macroscopic} as long as the radial confinement by the lattice laser remains negligible so that this relationship between superradiance gains is assumed to be independent on the lattice depth. 
The absolute gain for SR0 is given by $G_{ \vec k_{SR0}}=\frac{3 \Gamma |\Omega_0|^2}{8 k_S^2 \delta^2 W^2}$, with $\Gamma$ the excited state spontaneous emission rate,  so that $G_{ \vec k_{SR0}}=G_0 / \left(\int_{-d/2}^{d/2} d x w^4(x)\right)^{2/5}$ with $G_0$ the gain at 0 lattice depth, $\int_{-d/2}^{d/2} d x w^4(x)$ characterizing simply the compression along the $x$ axis due to the optical lattice. 
Exact correction is considered in our calculation \endnote{The calculation of the effective coupling constant is performed following the results of reference \cite{krämer2003bose}. Given the experimental uncertainties of our data, the integral of the Wannier function can be safely evaluated without taking into account interactions.}, though the harmonic oscillator approximation result $\int d x w^4(x) = \sqrt{\pi/2} (V_0/E_r)^{1/4}$ already good for $V_0>3 E_r$ implies $G_{ \vec k_{SR0}}$ varies only as $(V_0/E_r)^{-1/10}$ with the lattice depth \cite{krämer2002macroscopic}. 
In Eq. (\ref{rate}), $\hat{c}_0^{\dagger}\hat{c}_0$ and $\hat{c}_0\hat{c}_0$ were replaced by $N_0$ the number of atoms in the condensate at rest, that depends on time because of depletion during the superradiant process. 
We assume the operators $\hat{c}_q$ are well approximated by c-numbers of the form $c_q=\sqrt{N_{q}}e^{i \varphi_q}$, when the atom number $N_q$ of atoms in mode $q$ becomes large, $N_q>>1$, which is valid any time for the $\pm 2 \hbar \vec k_L$ modes if $V_0>0.2 E_r$. The initial conditions are then taken as $c_{\pm 2k_L}\left( 0 \right) = \sqrt{N_{\pm 2k_L}(0)}=\tilde{w}(-2 \vec k_L) \sqrt{N_t}$ with $N_t$ the total number of atoms. The operator standing for the interaction with vacuum was neglected in the dynamical equation of the $\pm 2 \vec k_L$ operators for this same reason. For SR0 at early times, we consider simply the mean number of scattered atoms in mode $k_{SR0}$ related to the operator $\hat{f}^\dagger_{ \vec k_{SR0}}$ and neglect fluctuations. As long as depletion of the condensate is not large, i.e. $N_0(0)-N_0(t)<0.1 N_0(0)$, this number is given  by $N_{k_{SR0}}(t)=\exp (G_{k_{SR0}} N_0 t)-1$ \cite{Moore}. At later times, when $N_{k_{SR0}}\sim 10 >>1$ typically obtained for $t>10 \mu$s for $V_0<5 E_r$, $\hat{f}^\dagger_{ \vec k_{SR0}}$ can be neglected in Eq. (\ref{rate}) and $\hat{c}_{k_{SR0}}$ replaced by a c-number. For too high lattice depths, depletion of the condensate could become non negligible during the first stage but we checked this is actually not the case considering our experimental conditions. 

The depletion of the 0 order leads to mode competition, each scattered photon corresponding to a scattered atom from the condensate.
Without initial lattice grating, maximum gain is obtained along the long axis of the BEC. As the lattice depth is increased, the seeding of the $-2 \hbar \vec k_L$ order, proportional to $\tilde{w}(-2 \vec k_L)$, increases and favors matter wave amplification because the exponential growth of the $-2 \hbar \vec k_L$ order reduces drastically the gain for superradiance along other modes.

\begin{figure}
[hptb]
\begin{center}
\includegraphics*[
width=6.5cm,height=4.0cm]{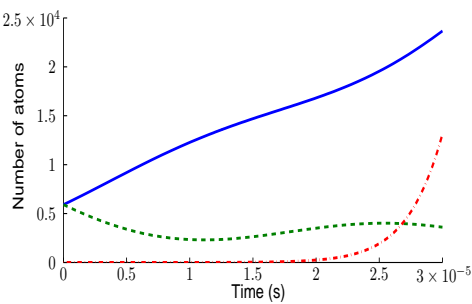}%
\caption{(Color online) Computed time evolution of the numbers of atoms in the $- \hbar 2 k_L$ mode or SRB mode (solid blue line), in the SR0 mode (dot-dashed red line) and in the $2 \hbar k_L$ mode (dashed green line). Conditions are the ones of Fig. \ref{threshold}, with a 3 $E_r$ optical lattice depth.}%
\label{Timeevolu}%
\end{center}
\end{figure}

Numerical computations of scattering are given in Fig. \ref{threshold} and show a qualitative agreement with the experiment when we assume $G_{0} = 2.9 ^{+ 0.4} _{-0.2}$~s$^{-1}$, which is of the order of our estimated theoretical value $G_{0}\approx $3.0~s$^{-1}$. 
The corresponding calculated time evolutions of the number of atoms in the different modes are shown in Fig. \ref{Timeevolu} for a 3~$E_r$ lattice depth. 
Since short superradiance pump pulses are used, our model neglects Doppler dephasing during the superradiance process, as it appears in Eq. (\ref{rate}) with a unique recoiling frequency $\omega_{ 2\vec k_L}$. The considered pump pulse durations were however long and weak enough for backward scattering to remain negligible in the SR0 process (cf. Fig. \ref{tof}). Spatial effects should probably be included to reproduce the asymmetry of the superradiant gain $G_{\vec k_{SR0}}$ observed experimentally \endnote{One concomitant article explaining this asymmetry and based on the semi-classical theory of reference \cite{zobay2006spatial} will be published.} and to explain the asymmetric depletion of the mode at rest while amplification of the SRB mode occurs. 

\begin{figure}
[hptb]
\begin{center}
\includegraphics*[
width=6.5cm,height=4.0cm]{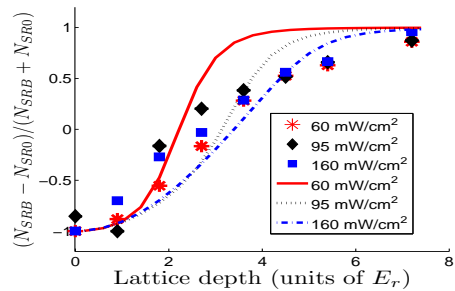}%
\caption{(Color online) Comparison of the number $N_{SR0}$ of atoms scattered in the forward SR0 mode (see text) and $N_{SRB}$ in the SRB mode, for different lattice depths and different intensities:  60 mW/cm$^2$  (stars), 95 mW/cm$^2$  (diamonds) and 160 mW/cm$^2$  (squares). The lines are a simulation of superradiance scattering for intensities, from top to bottom, equal to 60/cm$^2$  (solid line), 95/cm$^2$  (dotted line) and 160 mW/cm$^2$ (dot-dashed line), taking into account an additional loss rate $\gamma \sim 30/2\pi~$kHz of atoms in the mode at rest.}%
\label{thresholdpow}%
\end{center}
\end{figure}

To get a more precise understanding of the effect of quantum fluctuations, we studied mode competitions for different pumping intensities, i.e. different initial Rayleigh scattering rates and different superradiance gains.
In Fig. \ref{thresholdpow} is shown a study of the sensitivity of mode competition with the choice of pumping intensity, i.e. 60, 95 and 160 mW/cm$^2$.
As it is apparent on the data, the position of the threshold does not vary much with the choice of pumping intensity. At intermediate values of the optical lattice depth, a higher proportion of atoms scattered in the SR0 mode remains for lower pump intensities.

Such results are not well captured by our simple model. Indeed, our model predicts the variability of the position of the threshold should be very high.
To somewhat fit the experimental results, we added a loss rate on the number of atoms in the condensate at rest, proportional to the pump intensity  and equal to $\gamma \sim 30/2\pi~$kHz at 95~mW/cm$^2$ intensity, the loss effect being neglected for the other quasi modes in comparison to their growth rates.
The new choice of gain $G_{0}\approx $5.4~s$^{-1}$ at 95~mW/cm$^2$ intensity in Fig. \ref{thresholdpow} is however far from the theoretical value. Consequently, a model including a linear loss rate, though giving a better qualitative account of the behavior of competition in superradiance scattering, is insufficient in explaining accurately our results. 

\begin{figure}
[hptb]
\begin{center}
\includegraphics*[
width=6.5cm,height=4.0cm]{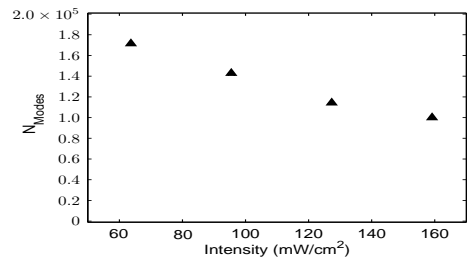}% 
\caption{Total number of atoms in different quasi modes N$_{\rm{Modes}}$ left after pumping (sum of all the quasi mode populations), for different pumping intensities.}%
\label{coherencedecrease}%
\end{center}
\end{figure}

In the experiment, it is actually found that $N_{\rm{Modes}}$, the total number of atoms populating the different quasi modes, decreases dramatically during pumping and this drop seems to depend on the pulse time duration, intensity and frequency detuning $\delta$. In Fig. \ref{tof}, this total number after pumping appears quite smaller than in the optical lattice. In order to get a better picture, the average number of such atoms versus pumping intensity is plotted in Fig \ref{coherencedecrease}. Its decrease is important, as much as 70$\%$ from 60~mW/cm$^2$ to 160~mW/cm$^2$. 
One first reason for the 70$\%$ decrease of $N_{\rm{Modes}}$ seems to correspond to a loss of coherence of the system occurring during the application of the pump pulse. Non cooperative Rayleigh scattering itself could cause only a limited $\sim 12\%$ drop of coherent atoms at 160~mW/cm$^2$ pump pulse intensity. Collisions when the atoms recoil is probably another source of decoherence. Obviously in the experiment, coherence of the sample is affected by the superradiance pulse but the measure of the degree of coherence is difficult because collisions, mainly in the form of s-wave scattering, are also present during the time of flight expansion. They form a strong background even without superradiance pumping and in that sense, absolute number measurements are difficult.
A smaller decrease, 30$\%$ from  60~mW/cm$^2$ to 160~mW/cm$^2$, also affects the total number of detected atoms in the ground $F=2$ state, calculated after integration over the whole imaging window, i.e. including the incoherent background of atoms.
This second loss could be related to pumping to the $F=1$ state, which we have not checked in the present experiment. Though it is neglected in our model, residual excitation to the $P_{3/2}$ state can result in spontaneous decay of atoms into the $S_{1/2},F=1$ state, with branching ratio 1/2 from the $F'=2$ excited state and 5/6 from the $F'=1$ state. In our Rayleigh scattering experiment with quantization axis determined by the Bias magnetic field along the BEC long axis, the main transition involved is $F=2,m_F=2\rightarrow F'=3,m_{F'}=3$ with $\sigma^+$ light involved. Nevertheless, Raman transitions are not completely canceled out and can lead to a loss of $F=2$ ground state atoms of about 3-4$\%$ at 160~mW/cm$^2$ intensity after 30 $\mu$s, calculated while neglecting cooperative scattering. Once considering cooperative scattering, the effects may be important and may have to be reconsidered in our model at high intensities.

In conclusion, mode competition in superradiant scattering from a Bose gas initially loaded in an optical lattice was studied for the first time. 
Superradiance with light scattering along end-fire modes was suppressed even at very low lattice depths because of the presence of the seeding of another superradiance mode by the optical lattice. 
Such results should provide a way to calibrate gains, thresholds or losses observed in superradiance scattering.
A simple model was developed to account for this mode competition.
Not surprisingly, we found this model is insufficient for charactering completely our experiment. We believe the present work will trigger new and interesting discussions. To the least, spatial effects, better analysis of decoherence processes or excitations and higher orders should all probably be included in the model \cite{müstecapliog,PhysRevA.71.033612,PhysRevA.72.041604,PhysRevA.75.033805}. A full quantum theory taking into account spatial propagation of the optical fields may be necessary \cite{PhysRevLett.105.220404,PhysRevA.82.023608}. 

This work received support from the National Fundamental
Research Program of China, grant No.
2011CB921501 and the National Natural Science Foundation
of China, grants No. 61027016, No. 61078026, No.
10874008 and No. 10934010.

\bibliographystyle{apsrev}
%\bibliography{SFMISRnoetal}

\end{document}